\documentclass[aps,prd]{revtex4}
\usepackage{graphics,psfrag}
\usepackage{graphicx,psfrag}
\usepackage{epic,eepic}
\usepackage{amsmath}
\usepackage{stmaryrd}
\usepackage{color} 
\usepackage{epsfig}
\usepackage{latexsym}
\usepackage{pstricks}
\newcommand{\be}{\begin{equation}}
\newcommand{\ee}{\end{equation}}
\newcommand{\bear}{\begin{eqnarray}}
\newcommand{\eear}{\end{eqnarray}}
\begin{document}

\title{Estimation of the large order behavior of the plaquette}
\author{Taekoon Lee}
\email{tlee@kunsan.ac.kr}
\affiliation{Department of Physics and Astronomy, 
Johns Hopkins University, Baltimore, MD 21218, USA and \\
Department of Physics, Kunsan National University, 
Kunsan 573-701, Korea}


\begin{abstract}
The universality of vacuum condensate can be exploited to 
relate the infrared renormalon caused
large order behaviors of different processes. As an application
the normalization constant of the large order behavior of
the average plaquette is estimated using the Adler function.

\end{abstract}

 

\maketitle

As is well known the perturbative expansion in weak coupling constant in
field theory is in general an asymptotic expansion,
with perturbative coefficients growing factorially at large orders.
There are two known sources for this  behavior. One is the factorial
growth of the number of Feynman diagrams at large order, which
may be understood using the instanton technique \cite{zinnjustin}.
The other is the renormalon  in which 
certain types of Feynman diagrams  give rise to the large order behavior 
via their infrared (IR) or ultraviolet (UV) behavior of 
Feynman integrals (For a review see \cite{beneke0}).
These renormalons cause singularities in Borel plane whose properties
can be studied by operator insertions for UV renormalons and operator
product expansion (OPE) for the IR renormalons \cite{parisi0,parisi1}. 
In quantum chromodynamics (QCD) the Borel summation
 of the asymptotic series by IR renormalon is inherently ambiguous, manifested by
 the presence of singularities on the integration contour. This ambiguity in
 Borel summation is supposed to be cancelled by the corresponding ambiguity in
 the vacuum condensates of the OPE. While this has not been proven there is 
 support for it from  two-dimensional nonlinear $\sigma$-models in 
 solvable large-N limit, where the ambiguity in imaginary part of the condensate
 is correlated with the contour choice of the Borel summation \cite{david0,david1}. 
Indeed, the nature of the renormalon singularities can be 
obtained via this cancellation
of the ambiguities \cite{mueller}.  
The purpose of our paper is to use this idea of ambiguity cancellation 
to relate the large order behaviors of
different processes.

 Consider a  real quantity $G(\alpha_s)$ that has an OPE expansion
 \bear
 G(\alpha_s)=C_0(\alpha_s)+C_1(\alpha_s) \langle O_1\rangle +\cdots
 \label{ope0}
 \eear
 where $C_0$ denotes perturbative contribution, $O_1$ is the operator for 
 the first power correction, and the suppressed are
  the higher dimensional operators. For simplicity, the dependence 
 on dimensional parameters in
 the Wilson coefficients and condensate are also suppressed.
 The Borel summation of the perturbative series is ambiguous,  which
  appears as contour dependent imaginary term that is to be cancelled by the
  ambiguity in the condensate $\langle O_1\rangle$.
  This means that
  \bear
  \frac{{\rm Im} C_0^{\rm BR}(\alpha_s)}{C_1(\alpha_s)}\,,
  \label{eq_ratio}
  \eear
  where $C_0^{\rm BR}$ denotes the Borel summed  of $C_0$,
  must be process independent, since the condensate, 
  being a vacuum property, should be universal,
   depending on no particular process. We note that when comparing
   (\ref{eq_ratio}) between two quantities
   the Wilson coefficients are to be computed in the same  
   renormalization scheme, unless the 
     condensate is scheme independent.
    Since the ambiguity is proportional to the normalization
  constant of large order behavior, this implies that the 
  large order behaviors of the quantities that have the OPE (\ref{ope0})
  with common condensate $\langle O_1\rangle $
  are all interrelated.
  To be specific, assume $C_0$ has perturbative expansion
  \bear
  C_0(\alpha_s)=\sum_{i=0} a_i\alpha_s^{i+1}\,.
  \eear
  This can be expressed in Borel integral as
  \bear
  C_0(\alpha_s)=\frac{1}{\beta_0}\int_0^\infty
   e^{-b/\beta_0\alpha_s} {\tilde G}(b) db
  \eear
  with the Borel transform given by
  \bear
  {\tilde G}(b)=\sum_{i=0} \frac{a_i}{i!} 
  \left(\frac{b}{\beta_0}\right)^i \,,
  \eear
  which is expected to have a finite radius of convergence, and
  $\beta_0$ is the one loop coefficient of the beta function
   given below in (\ref{betafunc}).
   The above mentioned cancellation of 
   ambiguities demands the Borel transform 
    have the singularity of the form
  \bear
  {\tilde G}(b)=\frac{\cal N}{(1-b/b_0)^{1+\nu}}(1+
  {\cal O}(1-b/b_0) )\,,
  \eear
  where $b_0$ is determined by the dimension of the operator and $\nu$
  by the renormalization group equation for the condensate and 
  are given as \cite{parisi1,mueller}
  \bear
  b_0=\frac{n}{2}\,, \quad
  \nu=\frac{n\beta_1}{2\beta_0^2}-\frac{\gamma_1}{\beta_0}
  \eear
  where $n$ is the dimension of $O_1$ and $\beta_i$ are the
  coefficients of the QCD beta function
  \bear
  \beta_{\rm QCD}(\alpha_s)=\mu^2\frac{d\alpha_s(\mu)}
  {d\mu^2}=-\beta_0\alpha_s^2 -\beta_1\alpha_s^3-\cdots\,,
  \label{betafunc}
  \eear
  and $\gamma_1$ is the coefficient at ${\cal O}(\alpha_s)$ of the anomalous
   dimension  of $O_1$.
  The large order behavior is determined by the singularity and is
  given by
  \bear
  a_i={\cal N} \frac{\Gamma(i+\nu+1)}{\Gamma(\nu+1)} 
   \left(\frac{\beta_0}{b_0}\right)^i (1+{\cal O}(1/i))\,.
  \eear

  The singularity causes the Borel integral depend on the choice of 
  the contour,  rendering the integral ambiguous. Taking the contour
  along the positive real axis on the upper half plane, the ambiguity, given by
 the imaginary part of the Borel integral
  \bear
  C_0^{\rm BR}(\alpha_s)=\frac{1}{\beta_0}\int_{0+i\varepsilon}^{\infty 
  +i\varepsilon} e^{-b/\beta_0\alpha_s} {\tilde G}(b) db\,,
  \eear
  where $\varepsilon$ denotes a positive infinitesimal,
 is obtained  as
  \bear
 {\rm Im} C_0^{\rm BR}(\alpha_s)= 
 {\cal N}\sin(\nu\pi)\Gamma(-\nu)(b_0/\beta_0)^{1+\nu}e^{-b_0/\beta_0\alpha_s}
 \alpha_s^{-\nu}
  (1+{\cal O}(\alpha_s))\,.
  \label{imag_part}
  \eear
This imaginary part is to be cancelled by that of the condensate, hence
  \bear
  {\rm Im} C_0^{\rm BR}(\alpha_s)+ C_1(\alpha_s) {\rm Im} 
  \langle O_1\rangle=0\,,
  \eear
 which means ${\rm Im}C_0^{\rm BR}(\alpha_s)/C_1(\alpha_s)$ is 
  process independent.
   Since the normalization is proportional to the
  ambiguity this allows one to interrelate normalizations 
  among different processes, and also shows that the normalization
  must be proportional to the leading order coefficient of the Wilson 
  coefficient $C_1$.
  
   As an application, let us consider 
   the average plaquette and the Adler function.
   Both have
  the gluon condensate 
\bear
  \langle G^2\rangle\equiv -\langle\frac{\beta_{\rm QCD}(\alpha_s)}
  {\pi\beta_0\alpha_s}G_{\mu\nu}^2\rangle
  \label{condensate}
\eear
as the leading operator for power correction, hence the large orders
of these can be related.
The OPE for the average plaquette $\text{U}_{\boxempty}$ is given by
\bear
P(\beta)\equiv\langle 1- \frac{1}{3}\text{Tr} \,\text{U}_{\boxempty}
\rangle=P_0(\alpha_{s\boxempty}) + 
Z(\alpha_{s\boxempty})\langle G^2 \rangle a^4 +O(a^6) \,,
\label{ope}
\eear
where
\bear
Z(\beta)=\frac{\pi^2}{36}(1 +{\cal O}(\alpha_{s\boxempty}))\,,
\eear
 $a$ is the lattice spacing, and 
$\alpha_{s\boxempty}=3/2\pi\beta$ denotes the bare coupling.
The OPE for the Adler function 
\bear
D(\alpha_s(Q))=-4\pi^2 Q^2 d\Pi(Q^2)/d Q^2-1\,,
\eear
where
\bear
\Pi(Q^2)=\frac{i}{3 Q^2} \int d^4x e^{iqx} 
\langle 0|T J_\mu(x) J^\mu(0)|0\rangle\,,
\eear
with  $Q^2=-q^2$ and $J^{\mu}$  a flavor 
nonsinglet vector (or axial) current,
 is given by
\bear
D(\alpha_s(Q))=D_0(\alpha_s(Q))+
D_4(\alpha_s(Q))\frac{\langle G^2\rangle}{Q^4} +{\cal O}(1/Q^6) 
\label{adler_ope}
\eear
where
\bear
 D_4(\alpha_s)=\frac{2\pi^2}{3}(1+{\cal O}(\alpha_s))\,.
\label{wilson_adler}
\eear

 Since we are
interested in QCD with no light quark flavors 
to compare with the average 
plaquette of pure Yang-Mills theory, we  assume
that the quarks composing the current are massive so
 that they do not contribute to
IR renormalon but still satisfy $ m_{\rm quark}^2\ll Q^2$ to 
make the OPE (\ref{adler_ope}) valid. In this limit, quark bubbles
should drop from the renormalon diagrams and the only quark 
lines are those contracting the currents.
 
Since $n=4$ and $\gamma_1=0$  for the gluon condensate (\ref{condensate})
 the renormalon singularity for the plaquette and the Adler function
 can be written, respectively, as
 \bear
  {\tilde P}(b)&\approx &\frac{\cal N_{\rm P}}{(1-b/2)^{1+\nu}} (1+{\cal O}(1-b/2))\,,\nonumber \\
  {\tilde D}(b)&\approx &\frac{\cal N_{\rm D}}{(1-b/2)^{1+\nu}}(1+{\cal O}(1-b/2))
  \label{ren_sings}
  \eear
 with
 \bear
 \nu&=&\frac{2\beta_1}{\beta_0^2}=\frac{204}{121}\,.
 \eear
 Now the ambiguity cancellation between the Borel summed perturbative contribution
  and the gluon condensate (\ref{condensate}), along with the renormalization 
 scheme independence of the gluon condensate by the trace anomaly \cite{traceanomaly}, gives
 \bear
 \frac{{\rm Im} P_0^{\rm BR}(\alpha_{s\boxempty})}{Z(\alpha_{s\boxempty})a^4}=
 \frac{Q^4{\rm Im} D_0^{\rm BR}(\alpha_{s}(Q))}{D_4(\alpha_{s}(Q))}\,.
\eear
Applying the formula (\ref{imag_part}) to the Borel integral with
the singularities (\ref{ren_sings}) we get 
\bear
\frac{{\cal N}_{\rm P}}{{\cal N}_{\rm D}}=\frac{1}{24}
(Qa)^4 \exp\left[-\frac{2}{\beta_0}(\frac{1}{\alpha_s(Q)}-
\frac{1}{\alpha_{s\boxempty}})\right](1+{\cal O}(\alpha_s))\,.
\label{cor1}
\eear

Using the relation between the lattice coupling and the $\overline{\rm MS}$
coupling at $N_f=0$, where $N_f$ denotes the number of light flavors, \cite{luscher}
\bear
\frac{1}{\alpha_s^{\overline{\rm MS}}(Q)}=\frac{1}{\alpha_{s\boxempty}}
+2\beta_0\ln(aQ) -4\pi t_1 +{\cal O}(\alpha_s)\,,
\label{cor2}
\eear
where 
\be
\beta_0=\frac{11}{4\pi}\,,\quad t_1=0.46820
\ee
we get
\bear
{\cal N}_{\rm P}=
\frac{e^{\frac{8\pi}{\beta_0} t_1}}{24}\,\,{\cal N}_{\rm D}^{\overline{\rm MS}} 
= 28703\,\, {\cal N}_{\rm D}^{\overline{\rm MS}}\,.
\label{ration}
\eear
Note that in obtaining this the higher order corrections in 
(\ref{cor1}) and (\ref{cor2}) can be safely ignored, since
 the ratio on the left hand side of (\ref{cor1}) is independent 
 of the strong coupling and so
 they must cancel out. Thus, (\ref{ration}) is  exact.

 We now turn to the computation of the normalization constants.
 The normalization constant of a renormalon
 can be computed using the scheme in \cite{lee0,lee1}, which exploits the singularity and analytic
 property of the Borel transform to compute the normalization using the usual perturbative
 expansion. The result is a convergent series expression of the normalization.
  The speed of convergence of the series depends on the quantity
 involved as well as the renormalization scheme. For instance this yields a 
 rapidly converging series for  the static interquark potential
or the heavy quark pole mass, rendering the normalization to be
evaluated accurately with the first few orders of perturbation \cite{pineda,lee2}. Recently,
 the estimation of the normalization was confirmed by numerical simulation \cite{bali0,bali1}.
     Numerically, the series
 for the normalization for the plaquette does not converge well at the
 orders known so far and so it cannot be obtained through the scheme.
 On the other hand, the scheme yields a  converging series for the Adler function.

 With the Borel  transform (\ref{ren_sings}) 
 the normalization  ${\cal N}_{\rm D}$ is given by
 \bear
 {\cal N}_{\rm D}= R(2)
 \eear
 where
 \be
 R(b)=\tilde D(b) (1-b/2)^{1+\nu}
 \ee
 To express $R(2)$ in  a convergent series form the singularity at $b=2$ must
 conformally be mapped so that it becomes the nearest singularity to the
 origin. Since the nearest singularity in $b$-plane is
 the UV renormalon at $b=-1$ we may use a mapping like
 \be
 z=\frac{b}{1+b}
 \ee
 which maps the singularity at $b=2$  to  one at $z_0=2/3$, which is 
 the nearest one
 on $z-$plane.
 On $z$-plane the normalization can be written as
 \bear
 {\cal N}_{\rm D}= R(b(z_0))
 = \sum_{i=0} r_i z_0^i
 \label{series}
 \eear
 where the series is now convergent. The coefficients $r_i$ can
 be computed from the perturbative series for $D_0$.
  
 $D_0$ in $\overline{\rm MS}$ scheme to five-loop is given as 
 \cite{chetyrkin0,chetyrkin1}
 \bear
 D_0(\alpha_s(Q))=  a_s +d_1 a_s^2+d_2 a_s^3+ d_3 a_s^4\,,
 \eear
 where $a_s=\alpha_s(Q)/\pi$ and, at $N_f=0$,
 \bear
  d_1=1.98571\,,\quad d_2=18.2427\,,\quad d_3=135.792\,.
 \eear
   The corresponding Borel transform is given as
  \bear
  \tilde D(b)=\frac{1}{\pi}\left[1 +
  d_1\left(\frac{b}{\pi\beta_0}\right)+\frac{d_2}{2!}
  \left(\frac{b}{\pi\beta_0}\right)^2
  +\frac{d_3}{3!}\left(\frac{b}{\pi\beta_0}\right)^3\right]
  \eear
  with which (\ref{series}) gives
 \bear
 {\cal N}_{\rm D}^{\overline{\rm MS}}
 =\frac{1}{\pi}(1-0.41393+0.08069+0.23598)=\frac{0.90274}{\pi}
 \label{power}
 \eear
 The series converges well up to four-loop order but jumps at five-loop.
 This jump is typical of the series for a singular function and just may reflect 
 the singular nature of $R(b(z))$.
 Note that $R(b(z))$ is still singular at $z=z_0$, for $\nu$ is a fractional
 number, but being bounded
 its series is guaranteed to converge at $z_0$; Nevertheless, the convergence
 can be bumpy, unlike for the series of a smooth function.
 It is also interesting to see the behavior of the normalization at differing $N_f$.
 For the first few nonzero flavors we have
 \be
  {\cal N}_{\rm D}^{\overline{\rm MS}}=\left\{\begin{array}{l}
                          (1-0.40840+0.01607+0.18119)/\pi \quad \text{for} \quad N_f=1\,, \\
                          (1-0.39613-0.06313+0.11121)/\pi \quad \text{for} \quad N_f=2 \,,\\
                          (1-0.37421-0.16118+0.01910)/\pi \quad \text{for} \quad N_f=3\,,
                          \end{array}
                          \right. 
 \ee
 which shows a better convergence at 
 increasing flavor numbers, a behavior that was already observed
 with Adler function of electromagnetic current \cite{lee0}. 
  
\begin{figure}[t]
\psfrag{NN}[][]{\fontsize{20}{20} ${\pi\mathcal N}_{\rm D}^{\overline{\rm MS}}(Q)$}
\psfrag{x}[c][c]{\fontsize{24}{24} $\mu/Q$}
\begin{center}
\includegraphics[angle=0,width=9cm]{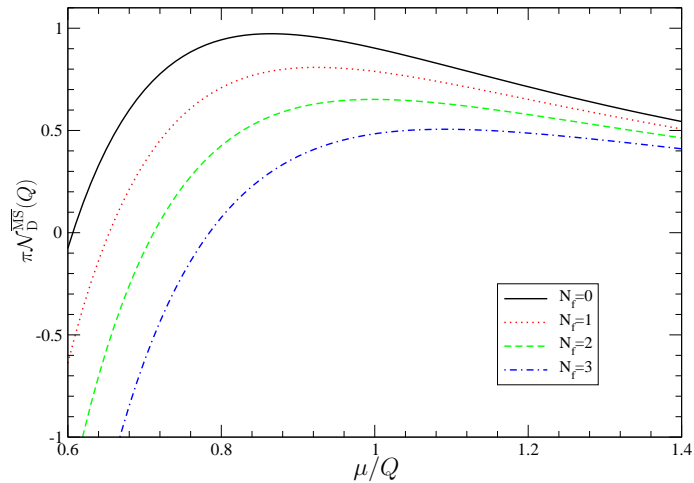} 
\end{center}
\caption{Scale dependence of the normalization 
constant at $\mu\!=\!Q$ of the Adler function.
}
\label{scaledep}
\end{figure}

  Now, taking the five-loop contribution 
 as the uncertainty in the estimate we conclude
  \bear
 {\cal N}_{\rm D}^{\overline{\rm MS}}=\frac{0.90\pm 0.24}{\pi}\,,
 \label{N_estimate}
 \eear
 from which we get the normalization for the average plaquette:
 \bear
 {\cal N}_{\rm P}= \frac{25833\pm 6889}{\pi}\,.
 \label{NN_estimate}
  \eear
  Considering the jump at five loop the uncertainty in this estimate can
  be too optimistic. To avoid such underestimation of the uncertainty we may
  also look at the renormalization scale dependence of the normalization constant.
  Being proportional to the gluon condensate the ratio 
  \bear
  \frac{{\rm Im} D_0^{\rm BR}(\alpha_{s}(\mu))}{D_4(\alpha_{s}(\mu))}
  \eear
 is scale independent, which means that the 
 normalization ${\cal N}_{\rm D}^{\overline{\rm MS}}(\mu)$ of the series in powers of
   $\alpha_s(\mu)$ of the Adler function scales as
   \bear
   {\cal N}_{\rm D}^{\overline{\rm MS}}(\mu)={\cal N}_{\rm D}^{\overline{\rm MS}}(Q) (\mu/Q)^4\,.
   \eear
  Thus the test of scale independence of 
  \bear
  {\cal N}_{\rm D}^{\overline{\rm MS}}(Q)={\cal N}_{\rm D}^{\overline{\rm MS}}(\mu) (Q/\mu)^4
   \eear
 can give a hint of the reliability of the estimate (\ref{N_estimate}).      
In Fig. \ref{scaledep} we see that  $Q$ is close to
the scale of minimal dependence, and considering the variation of the
normalization  about
 $\mu=Q$ it appears the error estimate in (\ref{N_estimate})
 is not unreasonable.      
 
 At this point it may be appropriate to isolate the universal portion of the
 normalization constants that is process-independent, by writing
 the normalization as 
 \bear
 {\cal N}^{\overline{\rm MS}}= c_0 {\cal N}_{G^2}^{\overline{\rm MS}}
 \eear
 where $c_0$ denotes the leading order coefficient of the process-depedent
  Wilson coefficient
 for the operator $G^2$. 
 From the estimate of the normalization for the Adler function and
 (\ref{wilson_adler}) we obtain
 \bear
 {\cal N}_{G^2}^{\overline{\rm MS}}=\frac{1.35\pm 0.36}{\pi^3}\,.
 \eear
 The process-dependence 
 of the normalization comes via $c_0$, a short-distance quantity,
 and ${\cal N}_{G^2}^{\overline{\rm MS}}$ is
  the process-independent part of the normalization,
   which may be regarded as the long-distance contribution 
   and an intrinsic
 property of the renormalon like 
 the strength $\nu$  or the position of the renormalon singularity.
 That the process-dependence comes only via a short-distance quantity
 should not be surprising, considering that in IR renormalon
 diagrams the large order behavior arises from bubble chains of
 arbitrarily long, far-infrared region; hence all process-dependence 
 should be a short-distance effect.
 
  Now note that  the normalization constant (\ref{NN_estimate}) is 
for the expansion in $\alpha_{s\boxempty}$. For the usual
power expansion in $1/\beta$
\bear
P(\beta)=\sum_{i=1} \frac{p_i}{\beta^i}
\eear
the large order behavior  is then given by
 \bear
 p_i=\frac{8\pi{\cal N}_{\rm P}\Gamma(i+\nu)}{11\Gamma(1+\nu)}
 \left(\frac{33}{16\pi^2}\right)^i (1+{\cal O}(1/i))\,.
 \label{lob}
 \eear

\begin{figure}[t]
\begin{center}
 \includegraphics[angle=0,width=10cm]
{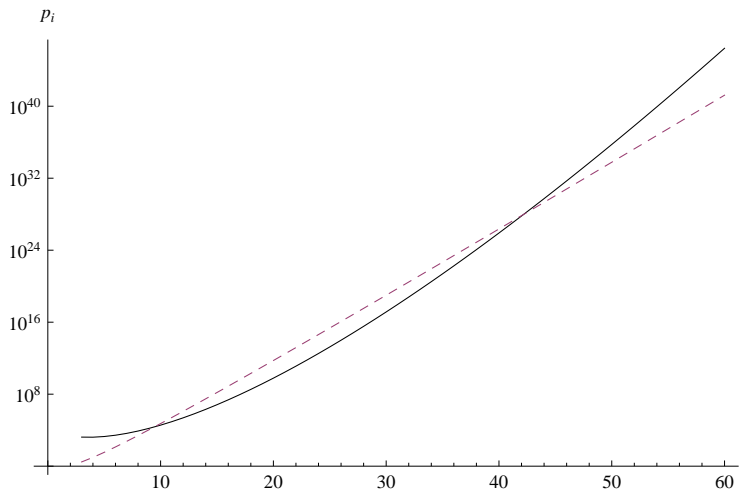}
\end{center}
\caption{Renormalon (solid) vs power law (dashed) behavior.}
\label{fig1}
\end{figure}

 The plaquette coefficients were computed in numerical 
 stochastic perturbation theory
 up to 20-loop orders \cite{direnzo0,direnzo1,horsley1}. 
 At these orders the coefficients grow much faster than 
 a renormalon behavior would suggest, and rather follow a power law.
 The plot (Fig.\ref{fig1}) of the renormalon behavior 
 (\ref{lob}) and power law \cite{horsley0}
 shows they meet at order  $i\sim 42$. This may suggest 
 the renormalon behavior would set in at orders around $i\sim40$.
  Recently, the renormalon behavior
 in heavy-quark pole mass was confirmed in numerical simulation of the
 coefficients to order $\alpha_s^{20}$ \cite{bali0,bali1}. 
 Our estimate of the large order behavior
 of the plaquette suggests a numerical evidence of 
 renormalon in plaquette would require
 much  higher order computations.  
  
\begin{acknowledgments}
The author is deeply grateful to the theory group in the 
department of physics and astronomy
at Johns Hopkins University and especially to Kirill
 Melnikov for hospitality.
This work was supported 
by the overseas research program of Kunsan National University.
\end{acknowledgments}

\bibliographystyle{apsrev}  
\bibliography{normalization}

\end{document}